\documentclass[twoside]{ahp}

\usepackage{joos_h}

\usepackage{graphicx}
\newcommand{\be}{\begin{equation}}
\newcommand{\ee}{\end{equation}}
\newcommand{\bea}{\begin{eqnarray}}
\newcommand{\eea}{\end{eqnarray}}
\newcommand{\ket}[1]{\left| #1 \right>}
\newcommand{\bra}[1]{\left< #1 \right|}
\newcommand{\scalar}[2]{\left< #1 | #2 \right>}	% scalar product
\newcommand{\lb}{\label}
\begin{document}

\title{Dynamical Consequences of Strong Entanglement}

\author{Erich Joos\\
{ Rosenweg 2, D-22869 Schenefeld}}

\maketitle

\begin{abstract}
The concept of motion in quantum theory is reviewed from a didactical point of view.
A unitary evolution according to a Schr\"odinger equation has  very
different properties compared to motion in classical physics. If the phase relations defining unitary dynamics are
destroyed or unavailable, motion becomes impossible (Zeno effect). The most important mechanism is dislocalization
 of phase relations (decoherence),
arising from coupling of a quantum system to its environment. Macroscopic systems are not frozen, although strong
decoherence is important to derive quasi-classical motion within the quantum framework. These two 
apparently conflicting consequences
of strong decoherence are analyzed and compared.
\end{abstract}

\section{Introduction}

It seems to be widely accepted by now that non-classical states of macroscopic objects can never show up in
the laboratory or elsewhere since they are unstable against decoherence. This explains superselection rules,
that is, kinematical restrictions in the space of all quantum states allowed by the superposition principle. The
observation that macroscopic objects are under ``continuous observation" by their natural environment paved the way for
our current understanding of the quantum-to-classical transition \cite{jo}.

Since in a consistent quantum treatment macro-objects are  obviously to be considered as open systems, their
dynamics can longer follow a Schr\"odinger equation. This alone invalidates  the textbook ``derivations" of the
classical limit via Ehrenfest theorems. Instead, one has to study the consequences of strong measurement-like
interaction of the considered system with its environment. The resulting entanglement not only superselects
certain states, which are then called ``classical" by definition, but also leads to dynamical consequences. Very
simple arguments seem to show that strong decoherence, that is, strong entanglement, leads
to slowing down of the dynamics of {\em any} system. However, the objects in our macroscopic world obviously
 {\em are moving}  and there seems
to be no ``Zeno effect". How this puzzle can be solved will be discussed in the
following sections.

\section{The quantum Zeno effect}

The quantum Zeno effect was discovered independently by several authors when studying the properties of decay
probabilities in quantum theory. The now popular term ``quantum Zeno effect" was introduced by Misra and
Sudarshan \cite{ms}.

Let a system be described by some ``undecayed"  state
$\ket{\Psi(0)}=\ket{u}$ at some initial time
$t=0$. The probability $P(t)$ to find it again in this ``undecayed" state at a later
time $t$ is
\be
P(t)=\left|\bra{u} e^{-i H t} \ket{u}\right|^2 \lb{eq1}
\ee
where $H$ is the Hamiltonian of the system. For small times we can expand $P(t)$,
yielding
\be
P(t)=1-(\Delta H)^2 t^2 + O(t^4) \label{equ2}
\ee
with
\be
(\Delta H)^2 = \bra{u} H^2 \ket{u} - \bra{u} H \ket{u}^2 .
\ee
The important feature to notice here is the {\em quadratic} time dependence of the
survival probability. This may be compared with the usual exponential decay law
\be
P(t)= {\rm exp}(-\Gamma t) ,\label{decay}
\ee
which leads to a {\em linear} time dependence for small times,
\be
P(t)=1 - \Gamma t + \dots \;   .
\ee

This raises the question, how these two differing results can be made
compatible. Both look fundamental, but they  obviously  contradict each other.
This conflict can be made even stronger, when we consider the case of repeated
measurements in a short time interval.

Suppose we repeat the measurement $N$ times during the interval $[0,t]$. Then the non-deacy (survival)
probability according to Equ. (\ref{equ2}) is
\be
P_N (t) \approx \left[ 1- (\Delta H)^2 \left(\frac{t}{N}\right)^2 \right]^N  > P(t) ,\label{equ6}
\ee
which for large $N$ gives
\be
P_N (t) =1 - (\Delta H)^2 \frac{t^2}{N}+ ... \stackrel{N \rightarrow \infty}{\longrightarrow} 1  .
\label{equ7}
\ee
This is the Zeno effect: Sufficiently dense measurements should halt any motion!

There is no Zeno effect if the system decays according to the exponential decay (\ref{decay})
law, since in this case trivially
\be
P_N (t) =\left( {\rm exp}\left(-\Gamma \frac{t}{N}\right)\right)^N 
= {\rm exp} (-\Gamma t) = P(t)  .
\ee
The conclusion is that any system showing a quadratic short-time behavior is very sensitive to measurements,
whereas an exponentially decaying system does not care about whether its decay status is measured or not, that
is, it behaves classically in this respect.

If a system is governed by the Schr\"odinger equation, as used in Equ.~(\ref{eq1}),
the transition probability for small times {\em must} start quadratically, hence
the exponential decay law can only be an approximation for larger times.
\footnote{There is a certain irony in this situation, since -- at least in popular
accounts -- exponential (``random") decay is used as a major argument that classical
physics has to be replaced by a new (quantum) theory. But there is no strictly exponential decay law in quantum
theory.}
 What
happens in the limit of ``continuous" observation? The Zeno argument seems to
show that there will be no motion at all!

To gain a better understanding of what is going on here,  I will discuss in the
following why motion is slowed down by measurements. In addition, the measurement
process itself will be described by a unitary evolution following the
Schr\"odinger equation as the fundamental law of motion for quantum states. It
will turn out, that the Zeno effect can be understood as a unitary dynamical process and
the collapse of the wave function is not required.

\section{Interference, Motion and Measurement in Quantum Theory}

\begin{figure}[htb]
\centering\includegraphics[width=0.3\textwidth]{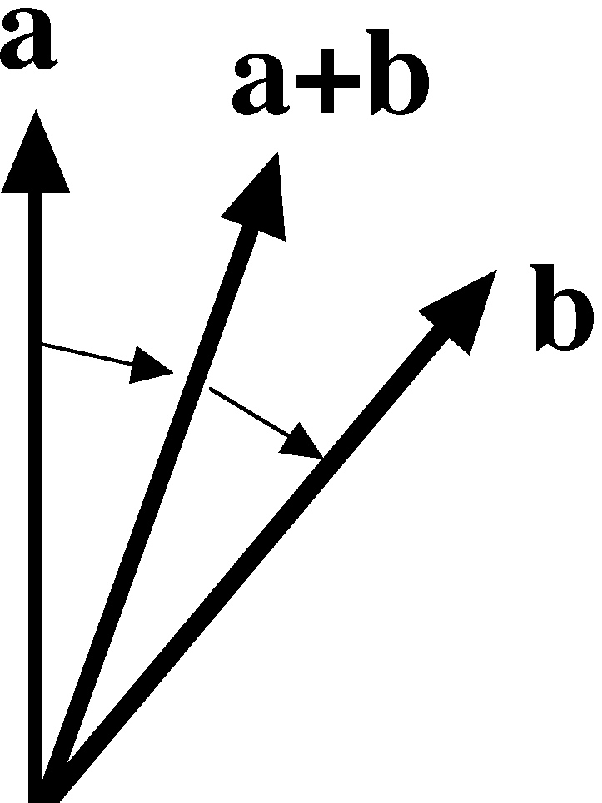}
\caption{Evolution in quantum theory can be viewed as a rotation connecting an initial
state $\ket{a}$ with a final state $\ket{b}$. For direct transitions, at intermediate times  superpositions
such as $\ket{a}+\ket{b}$ (neglecting normalization) are required for undisturbed motion.}
\label{figrot}
\end{figure}

Why does measurement slow down motion in quantum theory, but not in classical physics? The reason can be
traced back to the very nature of quantum  evolution. Quantum dynamics
is unitary and can be viewed as a rotation in Hilbert space, see Fig.~\ref{figrot}. 
If the Hamiltonian describes a direct unitary transition between two states $\ket{a}$ and $\ket{b}$,
the system has to go through a sequence of superposition states $\alpha(t) \ket{a} + \beta(t) \ket{b}$.
 An essential feature of such a
superpositions is the presence of interference (coherence). As is well known, such
a superposition has properties which none of its components has -- it is an
entirely new state. \footnote{This is the reason why stochastic models for quantum
evolution are unsuccessful: A superposition cannot be replaced by an ensemble of
its components.}
Unitary evolution from $\ket{a}$ to $\ket{b}$ requires all
the phase relations contained in the intermediate states $\alpha \ket{a} + \beta \ket{b}$. Phase
relations are destroyed by measurements, so it is not surprising that motion becomes
impossible in quantum theory if coherence is completely absent!

\begin{figure}[htb]
\centering\includegraphics[width=0.5\textwidth]{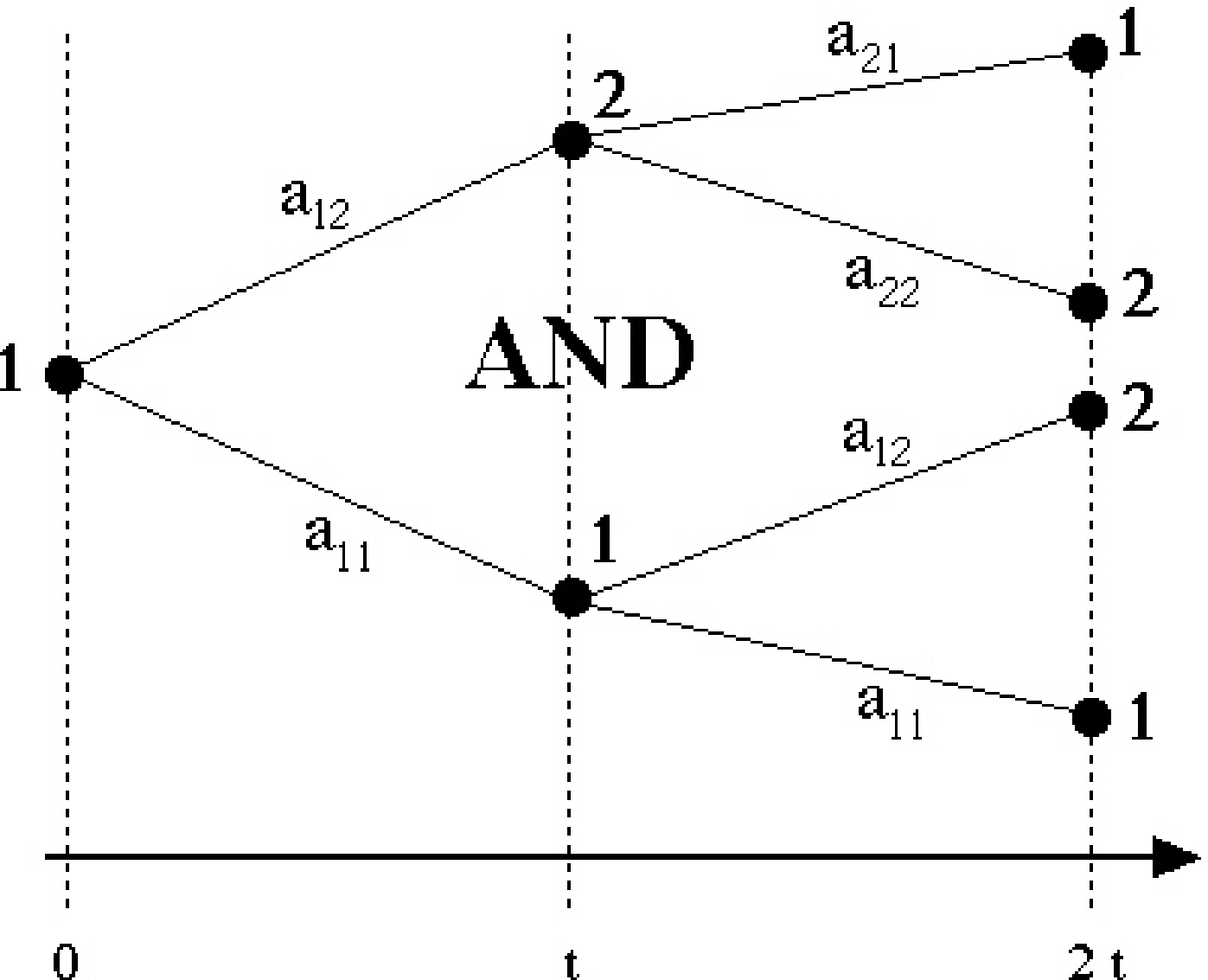}
\caption{Evolution of a two-state system away from initial state $\ket{1}$. The amplitude (and therefore the
probability) of state
$\ket{2}$ at time $2t$ depends on the phases contained in the superposition of $\ket{1}$ and
$\ket{2}$ at the intermediate time $t$, as in a double-slit experiment.}\label{figand}
\end{figure}

As an example consider the evolution of a two-state system from an initial state $\ket{1}$ as a two-step
process connecting times $0, t$, and $2t$, as shown in Fig.~\ref{figand}.
If $a_{ij}$ are transition amplitudes (calculated from the Schr\"odinger
equation) we have the chain
\bea
t=0: && \ket{1}\nonumber \\
&\longrightarrow &a_{11}\ket{1}+a_{12}\ket{2}\\
&\longrightarrow &
(a_{11}^2+a_{12}a_{21})\ket{1}+(a_{12}a_{22}+a_{11}a_{12})\ket{2}  .\nonumber
\eea
The final probability for state $\ket{2}$ at time $2t$ is then
\bea
P_2&=&\left|a_{12}a_{22}+a_{11}a_{12}\right|^2 
\eea
To study the Zeno effect we are interested in the behavior of $P_2$ for small times. In this limit it is given by
\be
P_2 \approx |V|^2 (2t)^2 \qquad \mbox{with} \qquad V=\bra{1}H\ket{2} .
\ee
Clearly the value for $P_2$ depends essentially on the presence of  interference terms. In a sense unitary
evolution is an ongoing double- (or multi-)slit experiment (without ever reaching the screen)!
\footnote{Obviously, the above model is nothing more than a very primitive version of the path-integral
formalism.}

Now compare this evolution with the same process, when a measurement is made at the intermediate time $t$. This
measurement may either be described by a collapse producing an ensemble (that is, resulting in $\ket{1}$ {\em or}
$\ket{2}$), or dynamically  by coupling to another degree of freedom. 
\begin{figure}[htb]
\centering\includegraphics[width=0.5\textwidth]{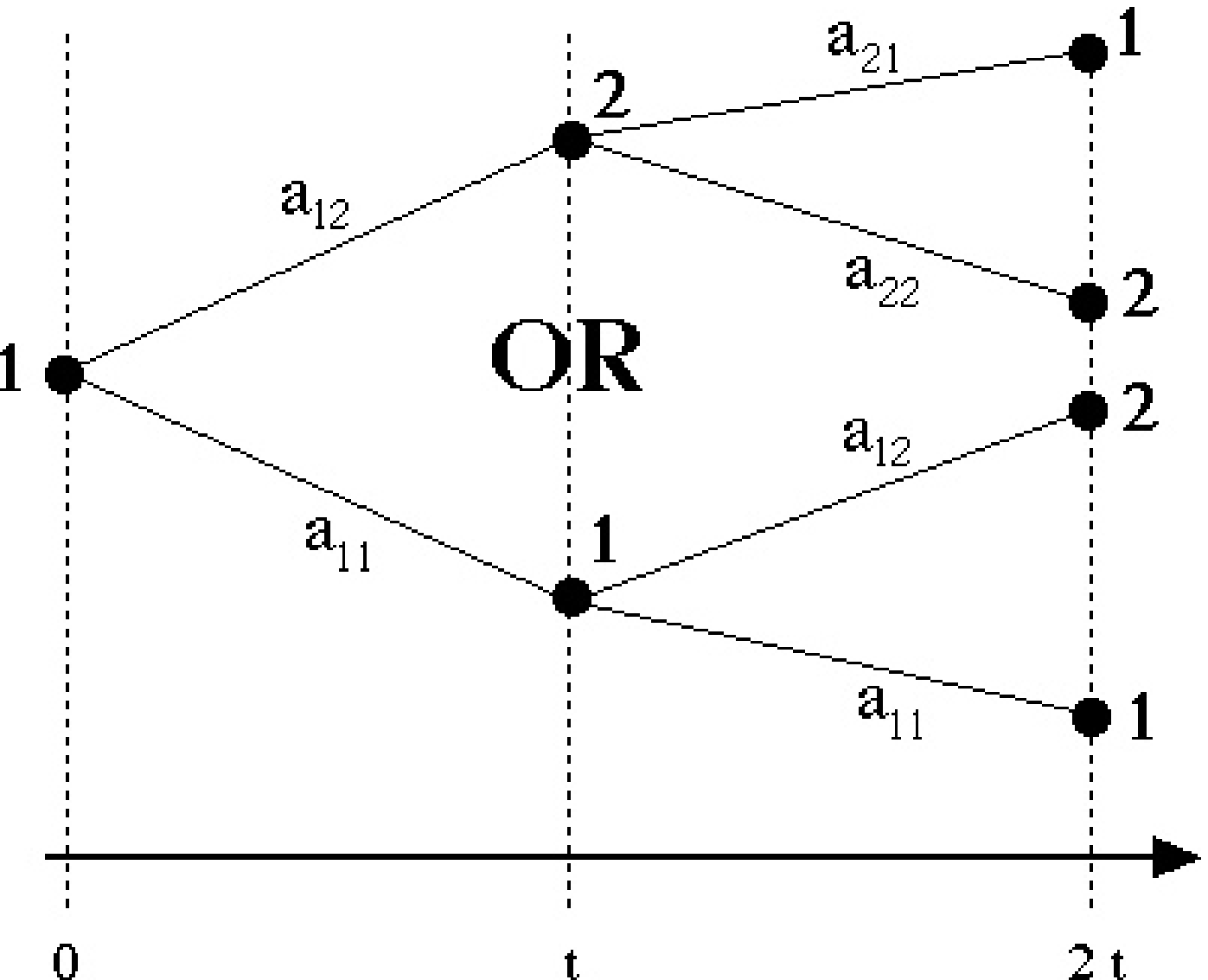}
\caption{Evolution of a two-state system with measurement. The probability for state $\ket{2}$ at time $2t$ results
solely from the transition probabilities to intermediate states at time $t$. The loss of phase
relations leads to a decrease of the total transition probability.}
\end{figure} 
In the latter case an entangled
state containing the system and the measuring device $\ket{\Phi}$ (or, generally, the system's
environment) ensues (more on this in the next section). The equations now look like
\bea
t=0: && \ket{1} \ket{\Phi} \nonumber\\
&\longrightarrow &(a_{11}\ket{1}+a_{12}\ket{2} )\ket{\Phi}\nonumber\\
&\longrightarrow &a_{11}\ket{1}\ket{\Phi_1}+a_{12}\ket{2} \ket{\Phi_2}\\
&\longrightarrow & (a_{11}^2\ket{\Phi_1}+a_{12}a_{21}\ket{\Phi_2})\ket{1}+
            (a_{12}a_{22}\ket{\Phi_1}+a_{11}a_{12}\ket{\Phi_2})\ket{2} \nonumber
\eea
(the third line describes the new measurement step) and the transition probability is given by
\bea
P_2&=&\left| a_{12}a_{22}\right|^2 +\left| a_{11}a_{12} \right|^2 \nonumber \\
&\approx& \frac{1}{2} |V|^2 (2t)^2 .
\eea

\noindent Since the interference terms are missing, we lose half of the probability! Clearly then, if we divide the
time interval not in two but into $N$ steps the transition probability is reduced by a factor
$1/N$:  the Zeno effect. This reduction is a sole consequence of entanglement
without any ``disturbance" of the measured system, since the measurement is assumed ideal in
this model. No coherence, no motion!

 The Zeno effect can also be seen more formally from the von Neumann equation for the density matrix. If
coherence is absent in a certain basis, the density matrix is diagonal, i.e.,
\be
\rho_{nm}=\rho_{nn}\delta_{nm} .
\ee
 But then no evolution is possible, since the von Neumann equation immediately yields
\be
\frac{d}{dt}\rho_{nn}=\sum_k \left( H_{nk}\rho_{kn}-\rho_{nk}H_{kn}\right) \equiv 0 .
\ee

\section{Measurement as a Dynamical Process: Decoherence}

To further analyze the Zeno effect I will consider a specific model for measurements of an $N$-state system.
As a preparation, let me shortly review the dynamical description of a measurement process.
In a dynamical description of measurement, the well-known loss of interference during measurement  follows from a
certain kind of interaction between a system and its environment.

Following von Neumann, consider an interaction between an $N$-state system and a ``measurement device" in the
form
\be 
\ket{n}\ket{\Phi_0} \longrightarrow {\rm exp}(-i H T) \ket{n}\ket{\Phi_0}=\ket{n}\ket{\Phi_n}\label{eqNeumann}
\ee
where $\ket{n}$ are the system states to be discriminated by the measurement device and $\ket{\Phi_n}$ are
``pointer states" telling which state of the system has been found. $H$ is an appropriate
interaction leading after the completion of the measurement (at time $T$) to orthogonal states of the
measuring device. Since in Equ.~(\ref{eqNeumann}) the system state is not changed, this
measurement is called ``ideal" (recoil-free). A general initial state of the system will -- via
the superposition principle -- lead to an entangled state,
\be
\left(\sum_n c_n\ket{n} \right) \ket{\Phi_0} \longrightarrow \sum_n c_n \ket{n}\ket{\Phi_n} .
\ee
This correlated state is still pure and does therefore {\em not} represent an ensemble of measurement results
(therefore such a model alone does not solve the measurement problem of quantum theory). The important point is that
the phase relations between different $n$ are {\em delocalized} into the larger system and are no longer available
at the system alone. Therefore the system {\em appears} to be in one of the states $\ket{n}$, formally described by
the diagonalization of its density matrix,
\bea
\rho&=&\sum_{n,m} c_m^* c_n \ket{n}\bra{m} \nonumber\\
&\longrightarrow& \sum_{n,m} c_m^* c_n \scalar{\Phi_m}{\Phi_n}  \ket{n}\bra{m} \\
&=& \sum_n |c_n|^2 \ket{n}\bra{n} \nonumber ,
\eea
where the last line is valid if the pointer (or environmental) states are orthogonal,
$\scalar{\Phi_m}{\Phi_n}=0$.

Any measurement-like interaction will therefore produce an apparent ensemble of system states. This
process is now usually called ``decoherence"\cite{jo}. Note that the origin of this effect is {\em not} a
disturbance of the system. Quite to the contrary: the system states $\ket{n}$ remain unchanged,
but they ``disturb" (change) the environment!

\section{Strong Decoherence of a Two-State System}

As a first application of the von-Neumann measurement model let us look at an explicit scheme for
a two-state system with Hamiltonian
\bea
H&=&H_0+H_{\rm int} \nonumber\\
&=& V(\ket{1}\bra{2}+\ket{2}\bra{1})+E\ket{2}\bra{2} \nonumber\\
&&+\gamma \hat{p} (\ket{1}\bra{1}-\ket{2}\bra{2}) .
\eea
The momentum operator $\hat{p}$ in $H_{\rm int}$ (last line) leads to a shift of a pointer wavefunction $\Phi(x)$ ``to the
right" or ``to the left", depending on the state of the measured system, $\gamma$ represents a measure of the strength
of this interaction. Because of the special structure of the Hamiltonian this interaction is recoil-free.
This model can be solved exactly and shows the expected damped oscillations. 
\begin{figure}[ht]
\centering\includegraphics[height=4.5cm]{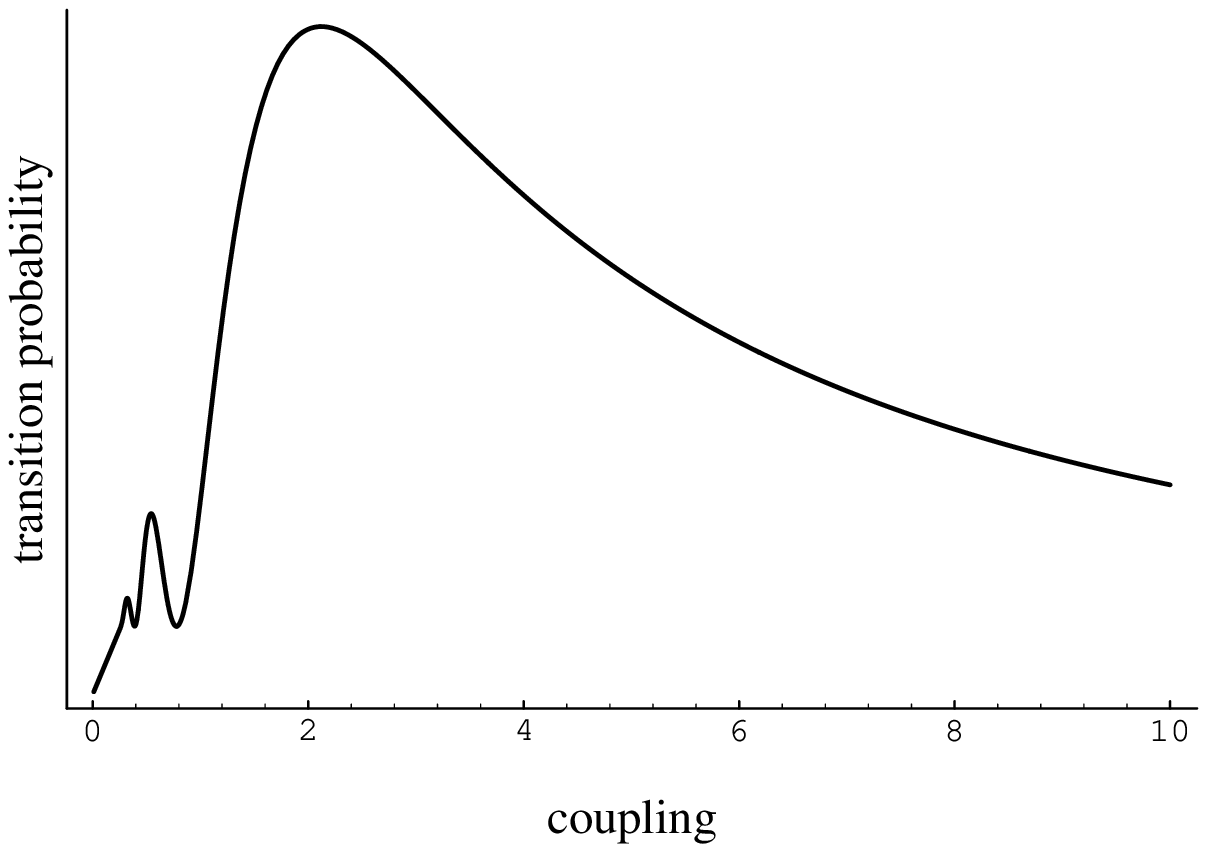}
\caption{Transition probability as a function of the coupling strength in a two-state model. For
strong coupling, transitions are always damped (Zeno effect).}\label{figzeno1}
\end{figure} 
In view of the Zeno effect we are
mostly interested in the limit of strong coupling. Here the solutions (calculated in perturbation theory) show two
interesting features, as displayed in Figs. \ref{figzeno1} and \ref{figzeno2} \cite{jo84}. First, the
transition probability from $\ket{1}$ to
$\ket{2}$ depends in a complicated way on the coupling strength, but for large coupling it always decreases with
increasing interaction. This is the expected Zeno behavior.

\begin{figure}[ht]
\centering\includegraphics[height=4.5cm]{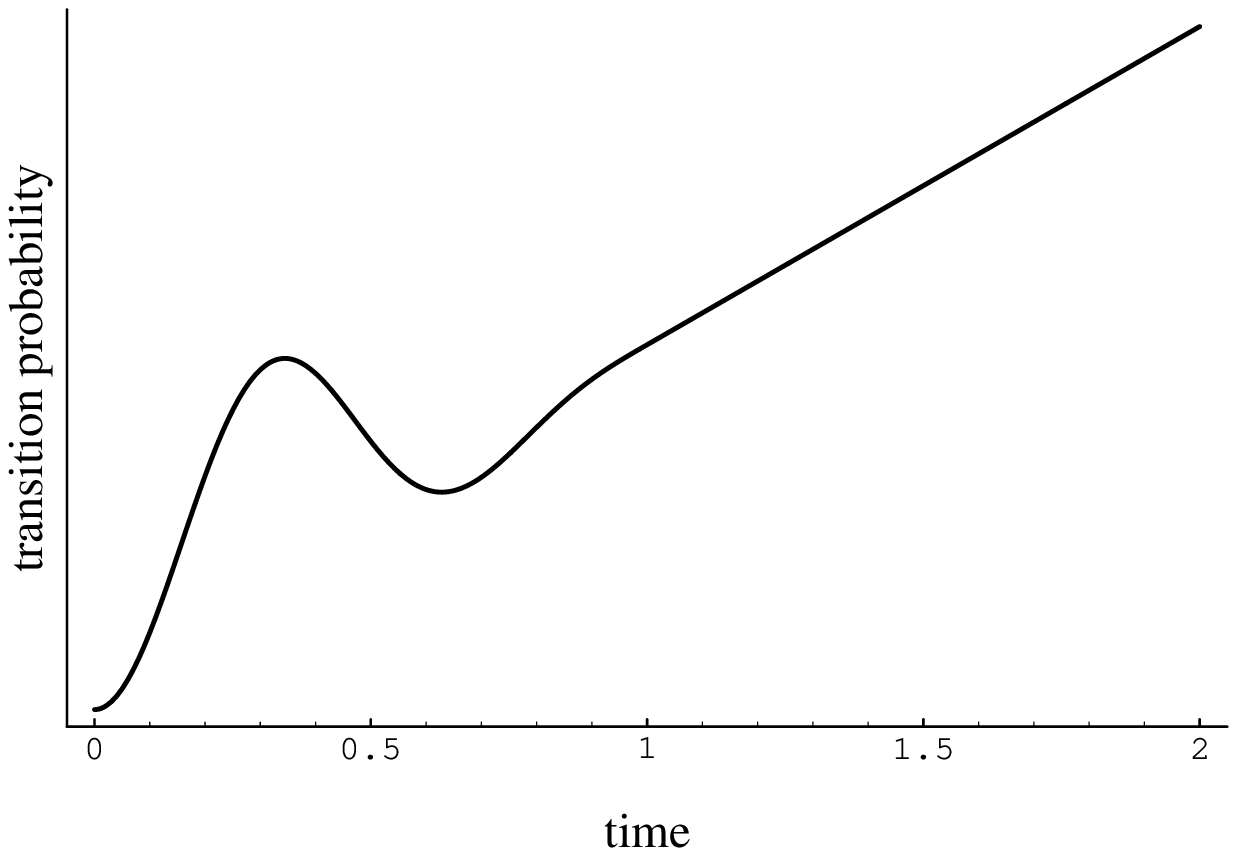}
\caption{Transition probability as a function of time. If the measurement can be
considered complete (here at $t \approx 1$), the transition probability grows
linearly (constant transition rates)}\lb{figzeno2}
\end{figure} 
If we look at the time dependence of the transition probability, we see the quadratic behavior for very
small times (as is required by the general theorem Equ.~(\ref{equ2})), but soon the transition
probability grows linearly, as in an exponentially decaying system (the rate, however, 
depends on the coupling strength).

A  realization of the quantum Zeno effect has been achieved in an experiment \cite{itano} where the two-state
system is represented in the form of an atomic transition, while the measurement process is realized by coupling
to a third state which emits fluorescence radiation, see. Fig.~\ref{figitano}.
\begin{figure}[ht]
\centering\includegraphics[trim=120 0 0 0,height=5.5cm]{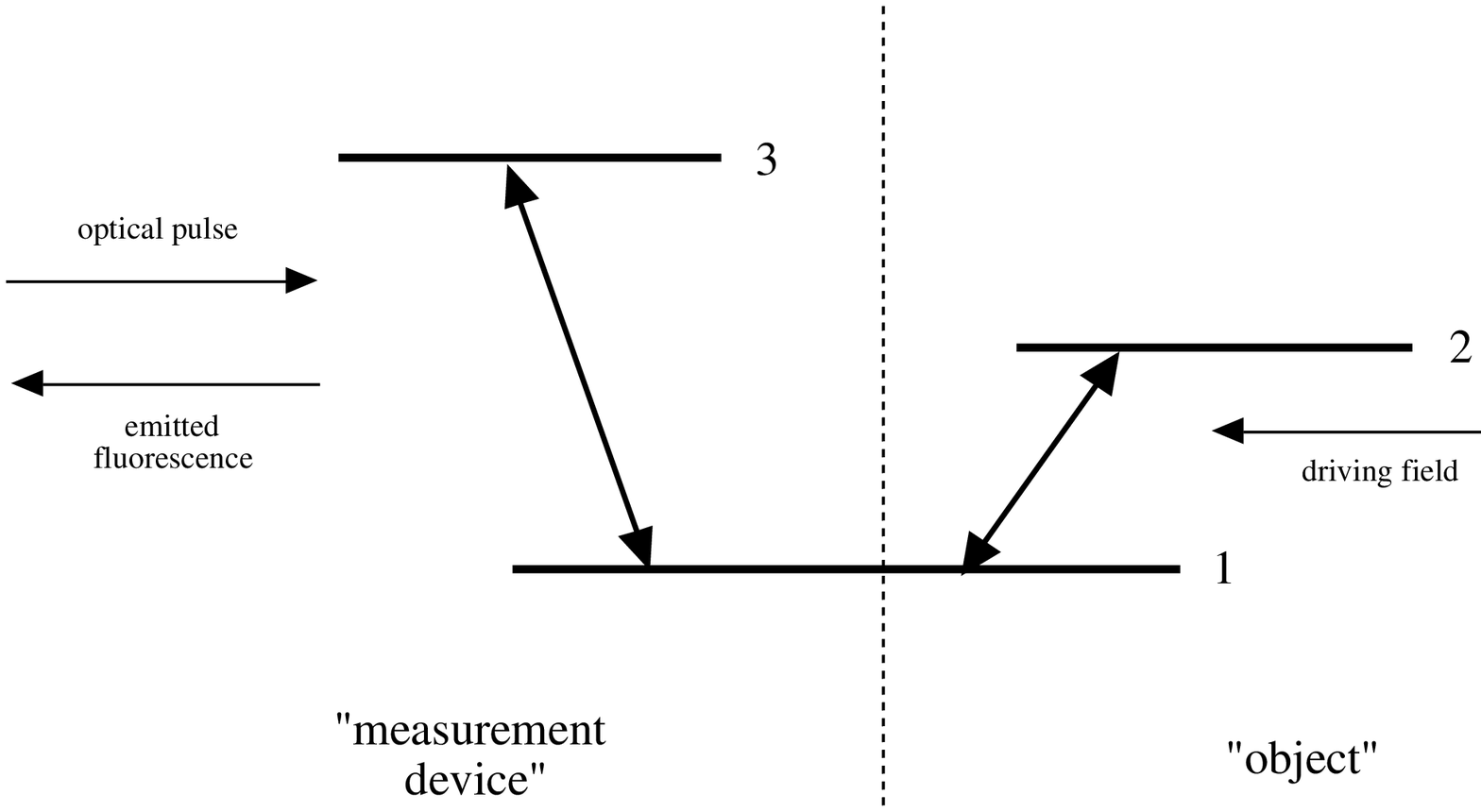}
\caption{Zeno experiment in atomic physics \cite{itano}. The two-state system under repeated observation is
represented by a transition between states $\ket{1}$ and $\ket{2}$. Measurement is accomplished
through an optical pulse leading to fluorescence from level $\ket{3}$ if the state $\ket{1}$ is
present.}\lb{figitano}
\end{figure} 

The Zeno effect also shows up in a curious way in a recent proposal of ``interaction-free measurement".
\begin{figure}[ht]
\centering\includegraphics[height=5cm]{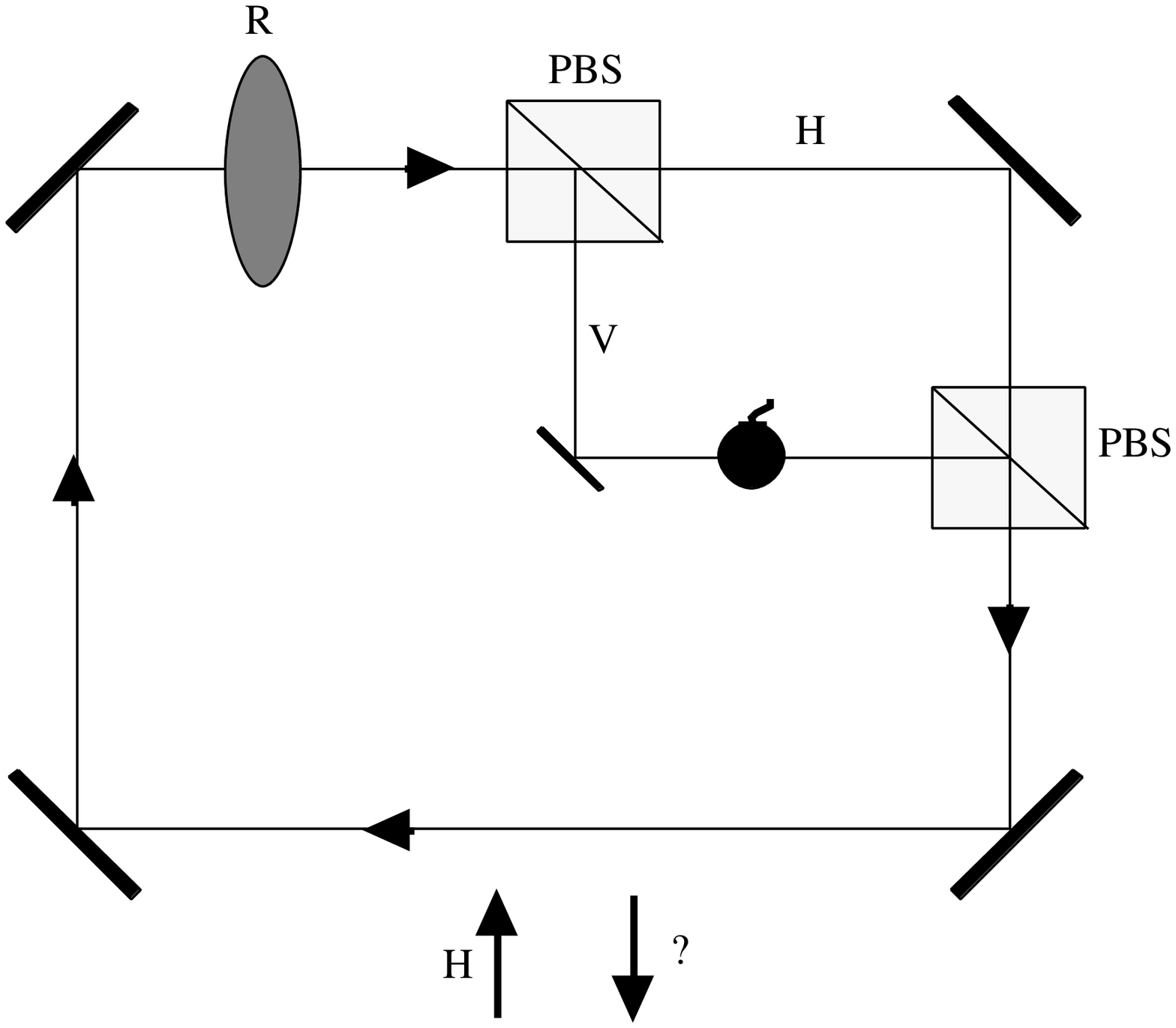}
\caption{Scheme of ``interaction-free interrogation" as a variant of the Zeno effect. Without the
absorbing object (the bomb), the polarization of the injected photon (initially horizontal) is
rotated by the rotator $R$ by a small angle on every passage. The two polarizing beam splitters $PBS$
have no effect, if properly adjusted, since horizontal and vertical components are recombined
coherently. If an absorbing object is present, the vertical polarization component is removed at every passage.
Inspecting the photon after many cycles allows one to infer the presence of the object with high
probability, while the photon is only very infrequently absorbed.}
\label{figIFM}
\end{figure} 

Early ideas about  ``negative result" or ``interaction-free"
measurements \cite{eli} can be combined with the Zeno mechanism \cite{ben}. One of these
schemes is exemplified in Fig.~\ref{figIFM}.
If a horizontally polarized photon is sent through $N$ polarization rotators (or
repeatedly through the same one), each of which rotates the polarization by an angle $\Delta \Theta =
\frac{\pi}{2 N}$, the photon ends up with vertical polarization. In this case  the probability to find
horizontal polarization would be zero,
\be P_H=0 .
\ee
If this evolution is interrupted by a horizontal polarizer (absorber) the probability of
transmission is (similar to Eqs. (\ref{equ6}) and (\ref{equ7})) given by
\be
P^{'}_H = \cos^{2N} \Delta \Theta = \cos ^{2N} \frac{\pi}{2 N} \approx 1 - \frac{\pi^2}{4 N}
\longrightarrow 1 .
\ee

To implement this idea, a photon is injected into the setup shown in Fig.~\ref{figIFM}
and goes
$N$ times through the rectangular path, as indicated. The initial polarization is rotated at R by an
angle
$\Delta
\Theta =
\frac{\pi}{2 N}$  on each passage. In the absence of the absorbing object, the polarizing beam
splitters, making up a Mach-Zehnder interferometer, are adjusted to have no effect. That is, the
 vertical component V is coherently recombined with the horizontal one (H)  at the second
beamsplitter to reproduce the rotated state of polarization. If, however, the ``bomb" is present,
the vertical component is absorbed at each step. After $N$ cycles, the photon is now still
horizontally polarized, thereby indicating the presence of the object with probability near one, or
has been absorbed (with arbitrarily small probability). For details of  experimental setups see
\cite{kwi}.

One should be aware of the fact that the term ``interaction-free" is seriously misleading since the
Zeno mechanism is a consequence of {\em strong} interaction. Part of this conceptual confusion is
related to the classical particle pictures often used in the interpretation of interference
experiments, in particular ``negative-result measurements".

\section{Strong Decoherence of  Many-State Systems}

Why does the Zeno effect not show up in our macroscopic world? I will consider two examples of classical
dynamics. The first is the motion of a massive object such as a dust particle or a planet. The second example
will be a reconsideration of Pauli's rate equation, describing classical random processes, where interference
apparently plays no role. In both cases it will turn out that (1) continuous measurement (i.e. decoherence) is
an essential ingredient for deriving classical motion and (2) the Zeno effect plays no role.

\subsection{Macroscopic objects}

With hindsight it seems to be a trivial observation that all macroscopic objects are strongly interacting with
their natural environments. The consequences have been analyzed only rather late in the history of quantum
theory  \cite{zeh,jz85}. One reason for this is certainly the prevailing Copenhagen orthodoxy. For generations
students were told that quantum theory should only be used for microscopic objects, while macroscopic bodies
are necessarily described by classical physics. 
\begin{figure}[ht]
\centering\includegraphics[width=0.7\textwidth]{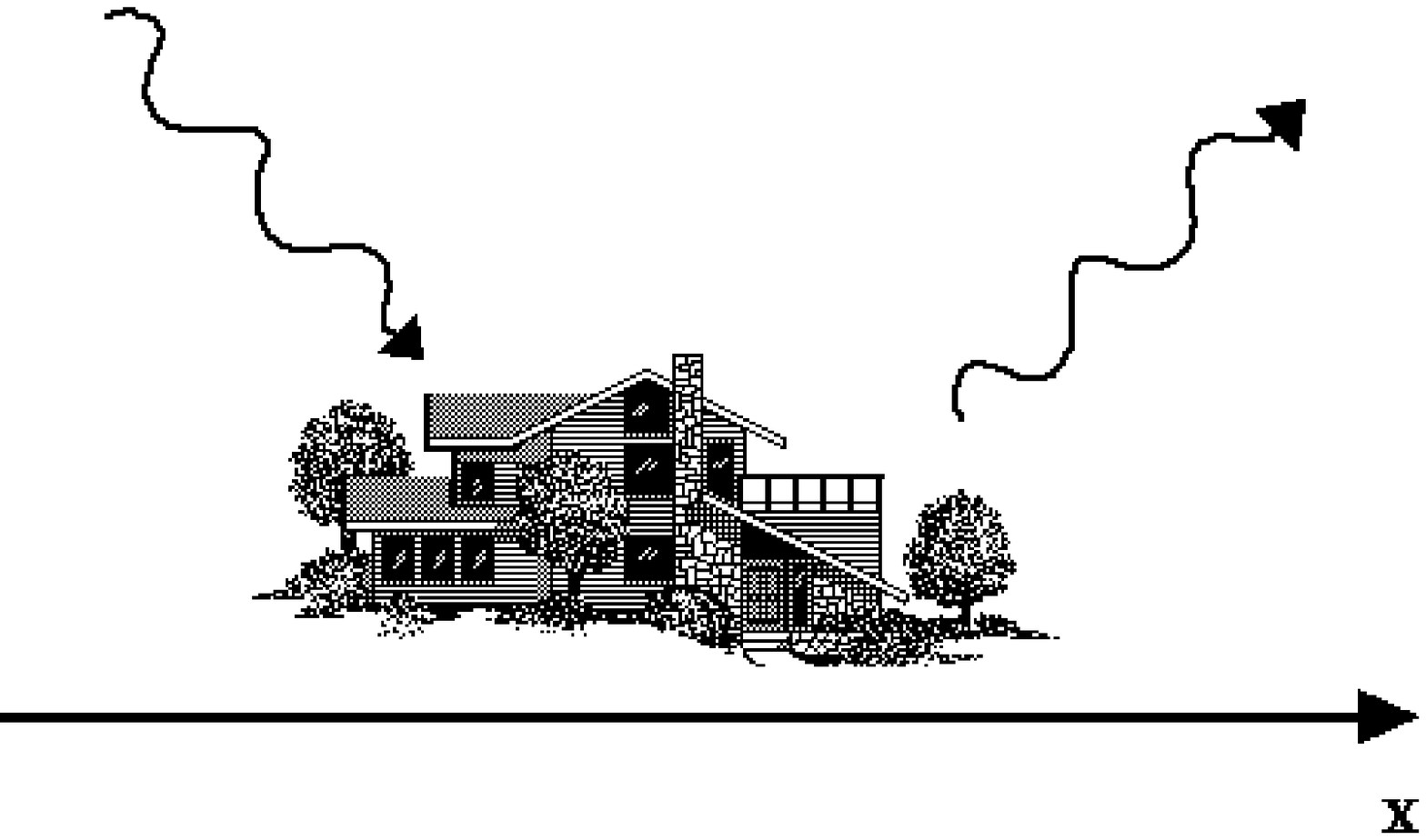}
\caption{Macroscopic objects can never be considered as isolated from their natural environment. Irreversible
scattering processes lead to ever-increasing entanglement.}
\end{figure}

The typical scenario is represented by scattering processes where the state of the scattered object, a
photon or a molecule, typically depends on the position of the macroscopic body. Quantitative estimates
\cite{jz85} show a strong effect, even in extreme situations, for example, a dust particle scattering only
cosmic background radiation.
For small distances, interference is damped according to
\be
\rho(x,x',t)=\rho(x,x',0) {\rm exp}[-\Lambda t (x-x')^2] \label{damp}
\ee
with a ``localization rate" $\Lambda$ given by
\be
\Lambda=\frac{k^2 Nv \sigma_{eff}}{V}\label{loc}
\ee
Here $k$ is the wave vector of the scattered particel, $Nv/V$ the incoming flux and $\sigma_{eff}$ of the
order of the total cross section. Some typical numbers are shown in the table.
\begin{table}
%%%%%%%%%%%%%
\caption{Localization rate $\Lambda$ in ${\rm cm}^{-2}{\rm s}^{-1}$
for three sizes of ``dust particles" and various types of
scattering processes  according to
(\ref{loc}).
This quantity measures how fast interference between different
positions disappears for distances smaller than the wavelength of the scattered particles, 
following Equ. (\ref{damp}). For
large distances, decoherence rates are just given by  scattering rates, and are thus independent of $x-x'$.}
\renewcommand{\arraystretch}{1.2}
\begin{tabular}{l|l|l|l}\noalign{\hrule height .8pt}
 & \hspace*{0.2cm}$a=10^{-3}\,{\rm cm}$  & \hspace*{0.2cm} $a=10^{-5}\,{\rm cm}$ & \hspace*{0.2cm} $a=10^{-6}\,{\rm cm}$ \\
 & \hspace*{0.2cm}dust particle\hspace*{0.2cm}& \hspace*{0.2cm}dust particle \hspace{0.2cm} &\hspace*{0.2cm}large molecule\\\hline
Cosmic background radiation\hspace{0.3cm} &   \hspace*{0.4cm}    $10^6$ &  \hspace*{0.4cm}	$10^{-6}$ &	   \hspace*{0.4cm}   $10^{-12}$ \\
300 K photons &  	 \hspace*{0.4cm}		 $10^{19}$ &	 \hspace*{0.4cm} $10^{12}$ &	    \hspace*{0.4cm}    $10^6$ \\
Sunlight (on earth) &  		  \hspace*{0.4cm}     $10^{21}$ & \hspace*{0.4cm}	$10^{17}$ &	  \hspace*{0.4cm}   $10^{13}$ \\
Air molecules & 			 \hspace*{0.4cm}      $10^{36}$ &     \hspace*{0.4cm}  $10^{32}$ &	   \hspace*{0.4cm}  $10^{30}$ \\
Laboratory vacuum&  	 \hspace*{0.4cm}	   $10^{23}$ &	   \hspace*{0.4cm}  $10^{19}$ & 	 \hspace*{0.4cm} $10^{17}$ \\
($10^6$ particles/$\rm{cm}^3$)&&&\\\noalign{\hrule height .8pt}
\end{tabular}
\renewcommand{\arraystretch}{1}
\end{table}
%%%%%%%%%%%%%%%%%%%%%%%%%%%%%%%%%%%%%%%%%%%%%%%%%%%%

The above equations are valid in the limit of small wavelengths, $k|x-x'|\ll 1$, comprising the effect of many
individually ineffective scatterng processes. The typical decoherence timescale according to
Equ. (\ref{damp}) is $t_{\rm dec} \approx \frac{1}{\Lambda |x-x'|^2}$.
In the opposite limit
$k|x-x'|\gg 1$, already a single scattering event destroys coherence. The decoherence timescale is then given by the
scattering rate (that is, $t_{\rm dec} \approx \frac{V}{N v \sigma_{\rm tot}} \approx \frac{k^2}{\Lambda}$).
A quantitative test of the quantum theory of spatial decoherence (\cite{jz85}, \cite{horn}   ) has been achieved in
interference experiments with large molecules \cite{hack}.

The equation of motion of, say, a dust particle,  is then no longer the von Neumann-Schr\"odinger equation, but
contains an additional scattering term (compare Equ. (\ref{damp}),
\be
i\frac{\partial\rho(x,x',t)}{\partial t}=\frac{1}{2m}\left(\frac{\partial^2}{\partial
x'^2}-\frac{\partial^2}{\partial x^2}\right)\rho -i\Lambda(x-x')^2 \rho .\label{CL}
\ee
If one analyzes the solutions of this equation, one finds that, for example, the Ehrenfest theorems for mean
position and momentum are still valid: The motion is {\em not} damped, although coherence between different
positions is destroyed. There is no Zeno effect.

The above equation of motion is a special case of more general equations which are studied under the topic ``Quantum Brownian Motion".
In addition to decoherence, these models include friction effects. A simple example is \cite{cal}
\bea
i\frac{\partial\rho(x,x',t)}{\partial
t}&=&\left[\frac{1}{2m}\left(\frac{\partial^2}{\partial x'^2}-\frac{\partial^2}{\partial x^2}\right)\right.
\nonumber\\ &&-i\Lambda(x-x')^2\nonumber\\
&&+\left. i\frac{\gamma}{2}(x-x')\left(\frac{\partial}{\partial x'}-\frac{\partial}{\partial x}\right)\right]\rho(x,x',t)
\eea
where
\be
\Lambda=m\gamma k_B T .
\ee
This model represents the environment as a bath of harmonic oscillators (with temperature $T$), coupled to
the mass point under consideration. The three
lines in Equ. (\ref{CL}) describe free motion, decoherence, and friction (damping constant $\gamma$), respectively.

In typical macroscopic situations, decoherence is much more important than friction. The ratio of decoherence to
relaxation rate can be estimated as
\be
\frac{\mbox{decoherence rate}}{\mbox{relaxation rate}}\approx m k_B T \left(\Delta x\right)^2 =
\left(\frac{\Delta x}{\lambda_{th}}\right)^2 ,
\ee
where $\lambda_{th}$ is the thermal deBroglie wavelength of the macroscopic body.
This ratio has the enormous value of about $10^{40}$ for a macroscopic situation (m=1 g, $\Delta x= 1$ cm)
\cite{zur}.

We can conclude from these models that
\begin{itemize}
\item
Newton's reversible laws of motion can be derived (to a very good approximation) from strong {\em irreversible}
decoherence.

\item 
The appearance of classical objects has its origin in low-entropy condition in the early universe and the
unique features of quantum nonlocality.

\item
Decoherence works {\em much} faster than friction in macroscopic situations.

\item 
Although coherence is strongly suppressed, no Zeno effect (slowing down of motion) appears.
\end{itemize}

\subsection{Rate equations}

The exponential decay law $P(t)={\rm exp}(-\lambda t)$ mentionend at the beginning is a special case of a general 
rate equation with transition rates $A_{\alpha \beta}$,
\be
\frac{d}{dt}P_\alpha=\sum_\beta A_{\alpha\beta} P_\beta=\sum_{\beta\ne\alpha}
\left( A_{\alpha\beta}P_\beta-A_{\beta\alpha}P_\alpha \right) .
\ee
Its quantum analogue, describing the dynamics of ``occupation probabilities" is usually called the ``Pauli equation",
\be
\frac{d}{dt}\rho_{\alpha\alpha}=\sum_\beta A_{\alpha\beta}\rho_{\beta\beta}\label{equpauli} .
\ee
An obvious feature of (\ref{equpauli}) is that interference terms do not play any dynamical role. On the other
hand, this cannot be true exactly, since then the von Neumann equation would lead to Zeno freezing,
\be
\frac{d}{dt}\rho_{\alpha\alpha}=\sum_\beta \left( H_{\alpha\beta} \rho_{\beta\alpha} - \rho_{\alpha\beta}H_{\beta\alpha}\right)\equiv 0 .
\ee
To further analyze these matters let us assume that the properties $\alpha$ in the rate equation are macroscopic in
the sense that they are {\em continuously observed by the environment}. The microscopic characterization is in the
following assumed to be given entirely by  energy, further degeneracies are neglected for simplicity. The
macroscopic feature $\alpha$ is measured by a ``pointer" as in the two-state Zeno model above, see
Fig.~\ref{paulimodel}. The Hamiltonian then reads \cite{jo84}
\bea
H&=& \sum_{\alpha E} E\ket{\alpha E}\bra{\alpha E} + \sum_{\alpha E \ne \alpha' E'}V_{\alpha E, \alpha' E'}\ket{\alpha E}\bra{\alpha'
E'}\nonumber\\
&&+\sum_{\alpha E} \gamma(\alpha) \hat{p} \ket{\alpha E}\bra{\alpha E}
\eea
As in the previous two-state model, the last line represents the (recoil-free) coupling to the ``pointer".

\begin{figure}[ht]
\centering\includegraphics[width=0.8\textwidth]{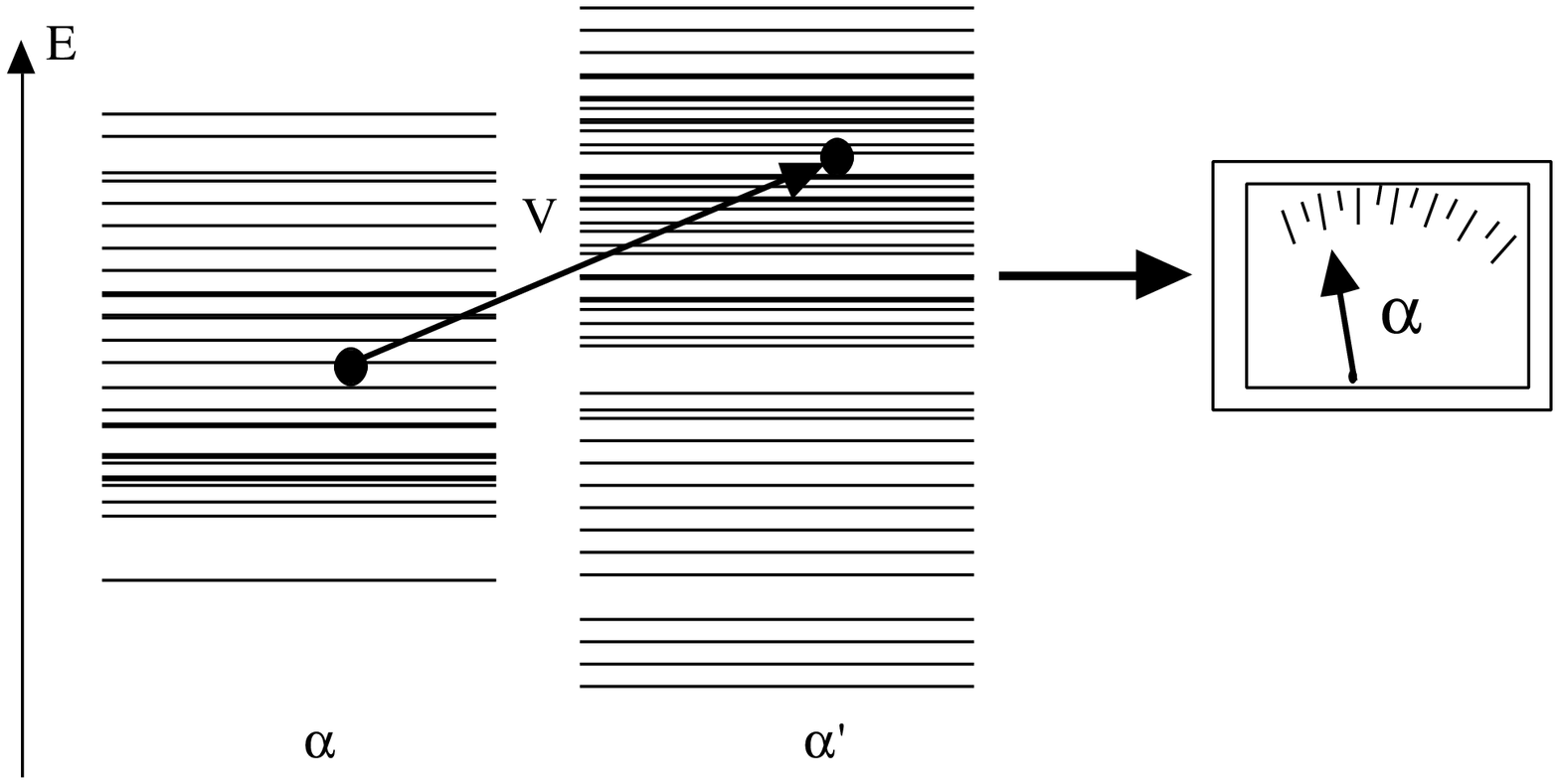}
\caption{Transitions between groups of states are monitored by a pointer. 
The symbolic measurement device in the figure represents the interaction with the
environment (which may or may not contain an experimental setup). Transition probabilities often
follow Fermi's Golden rule (rates governed by transition matrix elements $V$ and level densities at
resonance energy), but may be influenced by the presence of the environment monitoring certain
features
$\alpha$ of initial or final states.}\lb{paulimodel}
\end{figure}

Since we are interested in the limit of strong coupling to the pointer, we calculate the transition probability from property
$\alpha_0$ to another one, $\alpha$, in lowest order perturbation theory. Starting from
\be
\ket{\Psi(0)} =\ket{\alpha_0 E_0}\ket{\Phi} ,
\ee
where  $\Phi$ is the pointer state, the transition probability is
\be
P_{\alpha E}= 4 \int dp \left| V_{\alpha E, \alpha_0 E_0}\right|^2 \left| \Phi(p)\right|^2
\frac{\sin^2 (E-E_0 +\gamma(\alpha) p) t/2}{(E-E_0+\gamma(\alpha)p)^2} 
\ee
(assuming $\gamma(\alpha_0)=0$ for simplicity) .
This expression shows a familiar resonance factor, but now we have new resonances for each value of $p$ with weight $|\Phi(p)|^2$,
shifted from $E=E_0$ to a new value $E=E_0-\gamma(\alpha)p$. Summing over many states with property $\alpha$ gives
\be
P_\alpha \approx 2\pi t \int dE \frac{\sigma_\alpha (E) \left|V_{\alpha E,\alpha_0 E_0}\right|^2 }{\gamma(\alpha)}
\left| \Phi\left(\frac{E-E_0}{\gamma(\alpha)}\right)\right|^2
\ee
Three limiting cases can be extracted from this expression (see also Fig.~\ref{pauliresult}).
\begin{figure}[ht]
\centering\includegraphics[height=6cm]{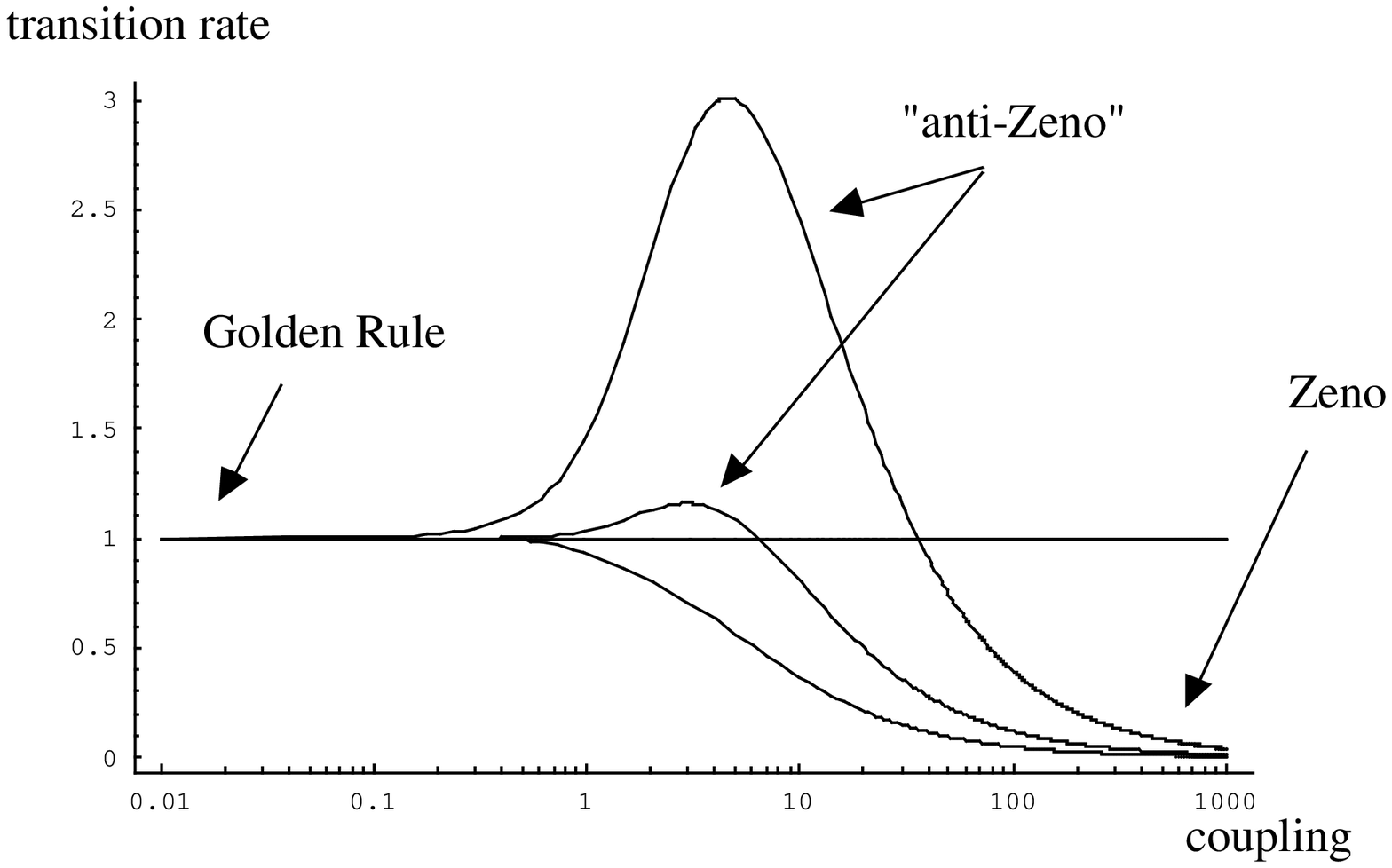}
\caption{Continuous coupling to a pointer changes the transition rate from an initial
state
$\ket{\alpha_0 E_0}$ to a group of final states in various ways. For small coupling we find the
standard Golden rule result (here normalized to unity). Increasing the coupling to the measuring
agent may in some cases increase the transition probability by shifting the effective resonance
frequency to regions with higher level density or larger transition matrix elements (anti-Zeno
effect). Strong interaction always leads to  decreasing  transition rates (Zeno effect).}
\lb{pauliresult}
\end{figure}
\begin{itemize}
\item Case 1: {\em Zeno limit}: For large coupling $\gamma(\alpha)$
we have
\be
P_\alpha \approx \frac{2\pi t}{\gamma(\alpha)} \int dE \; \sigma |V|^2(E) |\Phi(0)|^2 \sim
\frac{1}{\gamma(\alpha)}
\ee
Transitions are suppressed as expected.
\item Case 2: {\em Golden Rule limit}: For small coupling, transition rates become independent of  $\gamma$ and the
usual result is recovered,
\be
P_\alpha = 2\pi t \sigma_\alpha (E_0) |V(E_0)|^2 .
\ee
\item Case 3: {\em ``Anti-Zeno effect"}: If the contributions from each transition are comparable, that is, if
$\sigma |V|^2 \approx const.$ in the releveant interval $[E_{\rm min}, E_{\rm max}]$ then it is easy to see
that we have a smooth transition from the Zeno region to the Golden Rule limit. If this is not the case, it can
happen that in the intermediate range transition probabilities are {\em enhanced} above the Golden rule
value. This is occasionally called ``anti-Zeno effect".
\end{itemize}

\section{Summary}

We have seen that unitary evolution depends decisively on interference between
components of the wave function. If phase relations are lost, evolution is
hindered. This leads finally to the Zeno freezing of motion. No coherence, no
motion.

The destruction of phase relations can be understood as phase {\em
de-localization} arising from {\em unitary} quantum evolution, if the 
interaction of a system with its environment is taken into account. In this way,
the Zeno effect can be completely understood as a dynamical effect. No collapse
of the wave function is required, but only quantum nonlocality.

Many-state systems can escape Zeno freezing. This is important for the properties of
our experienced macroscopic world, but also for common ``quantum" features such
as radioactive decay, which happens whether or not a counter is setup to observe
the decay. (In fact, in most cases Nature herself provides the necessary
``counters".)

Systems with only a few degrees of freedom are very sensitive to quantum
entanglement and can therefore never escape the Zeno effect if they are interacting with other systems. Zeno
freezing can thus be used to delineate the borderline between classical and quantum objects.

\end{document}